\documentclass[conference]{IEEEtran}
\IEEEoverridecommandlockouts
\usepackage{cite}
\usepackage{amsmath,amssymb,amsfonts}
\usepackage{algorithmic}
\usepackage{amsmath}
\usepackage{graphicx}
\usepackage{textcomp}
\usepackage{xcolor}
\usepackage{booktabs}
\usepackage[colorlinks=true,bookmarks=false,citecolor=blue,urlcolor=blue]{hyperref} 
\usepackage{multirow}

\usepackage[pscoord]{eso-pic} 

\def\BibTeX{{\rm B\kern-.05em{\sc i\kern-.025em b}\kern-.08em
    T\kern-.1667em\lower.7ex\hbox{E}\kern-.125emX}}
\begin{document}

\title{Count-Min sketches for Telemetry: analysis of performance in P4 implementations}

\author{
\IEEEauthorblockN{J. A. Hern\'{a}ndez\IEEEauthorrefmark{1}, D. Scano\IEEEauthorrefmark{2}, F. Cugini\IEEEauthorrefmark{3}, G. Mart\'{i}nez\IEEEauthorrefmark{1}, N. Koneva\IEEEauthorrefmark{1}, A. S\'{a}nchez-Maci\'{a}n\IEEEauthorrefmark{1}, \'{O}. Gonz\'{a}lez de Dios\IEEEauthorrefmark{4} 
} 
\IEEEauthorblockA{\IEEEauthorrefmark{1}Dept. Ing. Telematica, Universidad Carlos III de Madrid, Spain.} 
\IEEEauthorblockA{\IEEEauthorrefmark{2}Scuola Superiore Sant'Anna, Italy}
\IEEEauthorblockA{\IEEEauthorrefmark{3}CNIT, Italy}
\IEEEauthorblockA{\IEEEauthorrefmark{4}Telefonica I+D Spain}
}



\maketitle

\begin{abstract}
Monitoring streams of packets at 100~Gb/s and beyond requires using compact and efficient hashing-techniques like HyperLogLog (HLL) or Count-Min Sketch (CMS). In this work, we evaluate the uses and applications of Count-Min Sketch for Metro Networks employing P4-based packet-optical nodes. We provide dimensioning rules for CMS at 100~Gb/s and 400~Gb/s and evaluate its performance in a real implementation testbed.
\end{abstract}


\section{Introduction}

Packet-optical nodes employing open operating systems like SONiC featuring high-speed long-reach coherent optical pluggables (e.g., 400G ZR+) have appeared in the optical arena to challenge classical transponder-based systems. 

In addition to having high-speed pluggables, such packet-optical boxes may include programmable ASICs leveraging P4 technology~\cite{P4_sigcomm}. P4 has demonstrated its potential in a wide range of scenarios, including advanced monitoring and telemetry, latency-aware scheduling and forwarding, 5G function acceleration, cyber-security, in-network AI, etc. A broad summary of these use cases can be found in~\cite{Filippo_overview_P4,Cugini_23_P4_invited}.

However, monitoring packet streams at line rates equal or above 100~Gb/s is extremely challenging, especially given the fact that flow size typically follow a Pareto  distribution. This means that a few minor flows (the elephants) contribute with the majority of packets and traffic in general and viceversa, most of the flows (the mice) contain only a few packets~\cite{elephant_flows,Dimensions_flows}. 

Indeed, the authors in~\cite{JURKIEWICZ202115} examine a large number of flows in a university campus in Polland, and show that almost two thirds of flows contain only 1 or 2 packets. This means that sampling packets randomly will very likely not observe a great portion of the packet flows, thus very inaccurate to identify flow cardinality or estimate heavy hitters. Well-known hash-based algorithms include: HyperLogLog (HLL) for estimating cardinality, Count-Min Sketch (CMS) for detecting heavy hitters, Bloom Filters (BF) for detecting whether a flow belongs to a set or not, etc. The authors in~\cite{namkung2022sketchlib} have implemented many of these hash-based algorithms in P4 and made the code open-source (SketchLib) for further experimentation by the research community.




This article briefly reviews the use cases and applications of CMS in the context of IP over Wavelength DIvision Multiplexing (IPoWDM) optical networks with P4-based packet-optical nodes. Section~\ref{sec:preliminars} briefly reviews CMS and network traffic characteristics. Section~\ref{sec:experiments} provides a number of simulation experiments and CMS design rule of thumbs for network scenarios at 100~Gb/s and some real experiments using Sketchlib on an Intel-based P4 Tofino scenario. Finally, section~\ref{sec:conclusions} concludes this work with its main findings and conclusions.

\section{Preliminars}
\label{sec:preliminars}

\subsection{Traffic statistics review}

In 2021, the authors in~\cite{JURKIEWICZ202115} released detailed statistics for the research community regarding packet traces collected at a university campus in Poland in year 2021~\cite{JURKIEWICZ202115}. Some of the results observed include: (1) average packet size of 870.6 Bytes, (2) Average flow size 68410 Bytes (78.5 packets), (3) Flows with only 1 packet: 47.8\%, and (4) Flows with only 1 or 2 packets: 65\%. Another important observation is that flow size follows a Pareto-like distribution, where the majority of flows contain very few packets, while very few heavy hitter flows comprise most of the traffic (in terms of packets), confirming past observations in different scenarios~\cite{elephant_flows,Dimensions_flows}.

Following these numbers, Table~\ref{tab:example_rates} summarises both flow and packet rates per switch port for a packet-optical node whose ports operate at line rates of 100~Gb/s and 400~Gb/s at different loads.

\begin{table}[htbp]
    \centering
    \begin{tabular}{|c|ccc|}
    \hline
        & & 100 Gb/s & \\
        & 10\% & 40\% & 70\% \\
        Flow rate & 18.25~K flow/s & 73~K flow/s & 127.75~K flow/s \\
        Packet rate & 1.54~M packet/s& 6.17~M packet/s & 10.8~M packet/s\\
    \hline    
    & & 400 Gb/s & \\
        & 10\% & 40\% & 70\% \\
        Flow rate & 73~K flow/s & 292~K flow/s & 511~K flow/s \\
        Packet rate & 6.16~M packet/s& 24.64~M packet/s & 43.12~M packet/s\\
    \hline
    \end{tabular}
    \caption{Flow and packet rate per port at 100 and 400~Gb/s}
    \label{tab:example_rates}
\end{table}


\subsection{Overview of CMS}



The CMS is a data-structure that allows to store the frequencies of each flowID in a compact manner~\cite{CORMODE200558}. It is very similar to having multiple Counting Bloom Filters. The challenge is again to store the frequencies of flows (especially the heavy hitters) where most of them are unique.

\begin{figure}[!htbp]
    \centering
    \includegraphics[width=0.8\columnwidth]{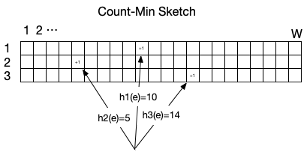}
    \caption{Example of a CMS structure, d=3 hash functions and W columns}
    \label{fig:cms_example}
\end{figure}

To do this, the CMS comprises a matrix with $d$ rows (one per hash function) and $w$ columns, as shown in Fig.~\ref{fig:cms_example}. When an element arrives (a packet in this scenario), the CMS computes all $d$ hash functions (one per row) and increases by one each position in the appropriate column. For instance, in the example of Fig.~\ref{fig:cms_example} with $d=3$ rows and $w=20$ columns, let us assume that the hashing of packet $e$ gives the  following results: $h_1(e)=10$, $h_2(e)=5$ and $h_3(e)=14$. These values imply that the positions $[1,10]$, $[2,5]$ and $[3,14]$ in the CMS matrix should be increased by one, as shown in the figure. Thus, every single packet traversing the port requires computing $d$ different hash functions of it and increasing the corresponding positions $[d,h_d(e)]$ of the CMS sketch.

After all elements (packets) are introduced, the CMS can be queried to get an approximate of frequency for a particular element. Consider we want to estimate the popularity of element $y$. To do this, the procedure is to hash element $y$ and take the minimum value. Consider for example, that the CMS array returns the following results: $h_1(y) = 6$, $h_2(y) = 7$ and $h_3(y) = 2$, and looking at positions 
$[1,6]$, $[2,7]$ and $[3,2]$ of the CMS matrix, we observe values 5, 7, and 5 again. The number of packets arrived so far with flowID = $y$ would then be $\min(5,7,5) = 5$ packets. This is an estimate and some errors may have occurred due to hash collisions (as it is probably the second counter $[2,7]$ since the other two counters output 5 packets). For this reason, it is necessary to accurately dimension the CMS structure for a given expected number of elements, that is, sufficient rows and columns.

In general, CMS is an $(\epsilon,\delta)$ probabilistic data structure, where the result returned is bounded to at most $\epsilon \| Count \|$ with probability $1-\delta$, while both $\epsilon$ and $\delta$ can be obtained from the CMS design parameters $d$ and $w$ as it follows:
\begin{equation}
    d = \left\lceil \ln(1/\delta) \right\rceil \quad \text{and} \quad w = \left\lceil \frac{e}{\epsilon} \right\rceil
\end{equation}

where $e$ is the estimation error.





\section{Scenario and experiments}
\label{sec:experiments}

\subsection{Network setting and CMS dimensioning}

Let us consider a monitoring time window of 0.1~seconds, where a blank CMS is zeroed at time $t_0=0$ and is populated with the number of packet arrivals until $t_{w}=0.1~s$. Following Table~\ref{tab:example_rates} (100~Gb/s, 40\% load), we should expect around 7,000 different flows whose size follows a zipf distribution. In this regard, we have simulated a packet trace with $N=7,000$ different flow IDs and packet sizes following a zipf distribution with scale $\alpha=1.1$ (following the examples of~\cite{jahernandez_ftth}). This means that the $k$-th flow has a frequency $f_k$ following:
\begin{equation}
    f_k = \frac{\frac{1}{k^\alpha}}{\sum_{n=1}^N \frac{1}{n^\alpha}},\quad k=1,\ldots,N
\end{equation}

\begin{figure}
    \centering
    \includegraphics[width=0.9\columnwidth]{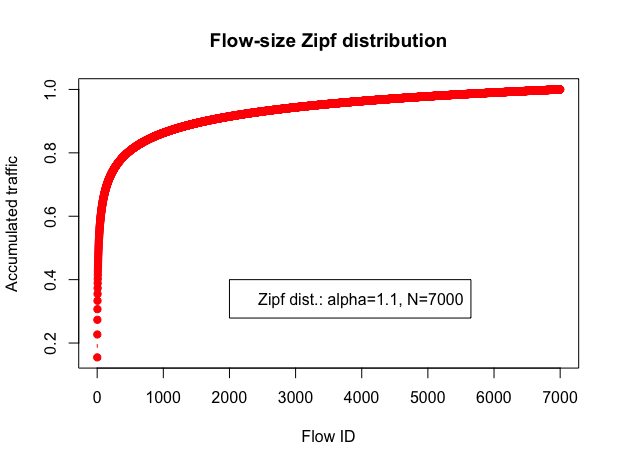}
    \caption{Flow-size distribution: Zipf-like CDF}
    \label{fig:zipf}
\end{figure}

As shown in Fig.~\ref{fig:zipf}, the largest flow contributes with 15\% of the total traffic while the top-20 heaviest hitters comprises 49\% of the total traffic (almost the same as the other 6,980 flows). In this scenario, identifying the top-20 flows allows operators to take decisions regarding their treatment, i.e. using different optical bands or alternative routes for load balancing and QoS guarantees.

\begin{figure}[!htbp]
    \centering
    \includegraphics[width=\columnwidth]{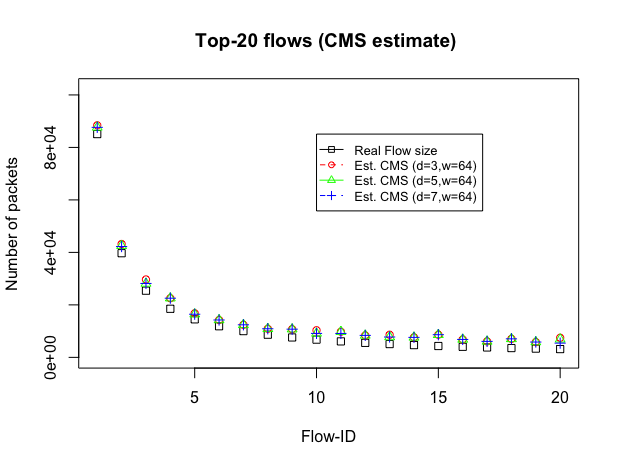}
    \includegraphics[width=\columnwidth]{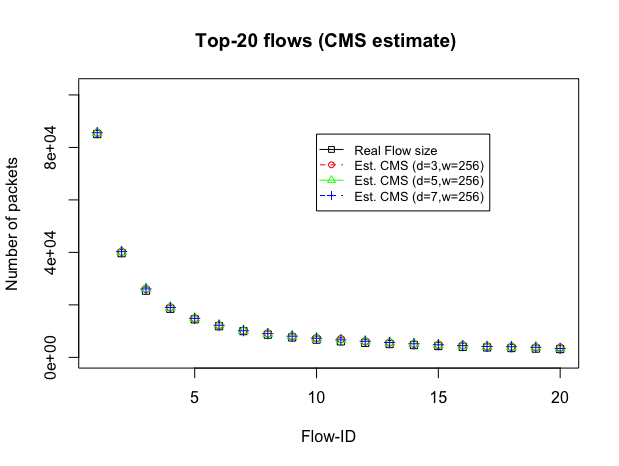}
    \caption{CMS accuracy at estimating top-20 heavy-hitter flows for different configurations of CMS sketch: (top) $d=\{3,5,7\}$ hashes and $w=64$ columns, and (bottom) $d=\{3,5,7\}$ hashes and $w=256$ columns}
    \label{fig:CMSscenarios}
\end{figure}

Fig.~\ref{fig:CMSscenarios} further shows the estimated flow size (in no. of packets) for the top-20 heaviest hitters for different CMS configurations, i.e. values of $d\in\{3,5,7\}$ hashes and $w=\in\{64,256\}$ columns (simulation code available in Github~\cite{github_cms_jahernandez}).

\subsection{Experimental validation}

\begin{figure}[!htbp]
    \centering
    \includegraphics[width=0.85\columnwidth, angle=270]{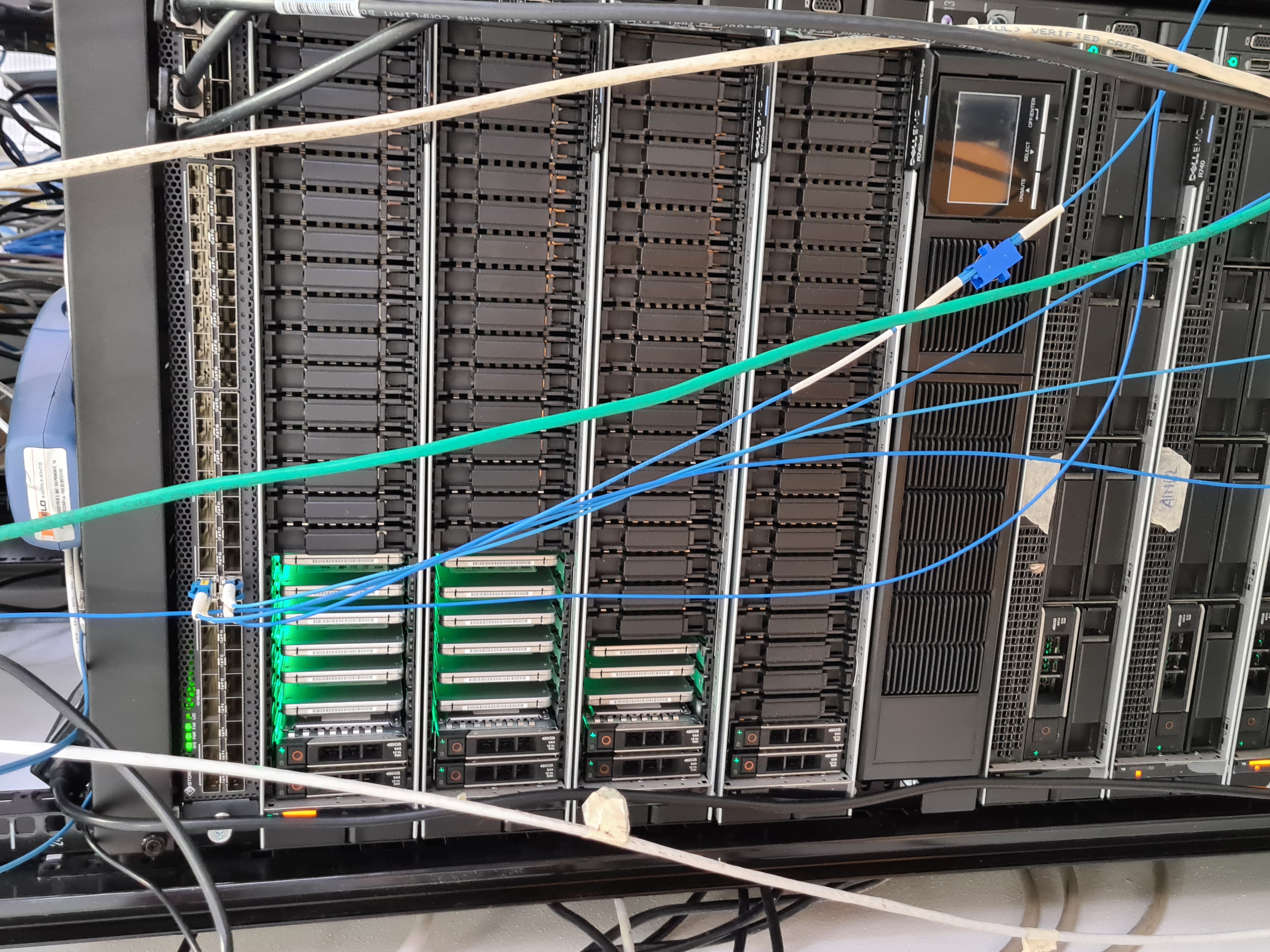}
    \caption{Experimental setup in lab}
    \label{fig:testbed}
\end{figure}

The CMS-based solution has been validated using a network switch, equipped with P4 ASIC, designed for data center operations and potentially suitable for IPoWDM implementations (as shown in Fig.~\ref{fig:testbed}). 

The traffic flows have been generated using a Spirent N4U. The P4 implementation uses five hash functions on the IP source address, each one producing a result between 0 and 4,096 bit ($=2^{12}$), while the total CMS occupies one register of size 65,536 bit.


In a first experiment, with the aim of collecting baseline performance, the traffic is sent from one source to one destination host, at a maximum rate of 9.85 Gb/s over a 10G port with fixed packet size of 512 bytes. The experienced latency to traverse the switch (including hashing and CMS processing) is $0.95~\mu s$.



Then, traffic is generated by 5 different sub-networks each one comprising 256 hosts (i.e 1,280 IP source addresses) that send traffic to another destination sub-network comprising 256 hosts. The average latency to traverse the switch interconnecting both source and destination subnetworks is again $0.95~\mu s$. This values slightly increases to $1.35~\mu s$ for packet size values of 1,024 bytes.




\section{Conclusions}
\label{sec:conclusions}

This work has evaluated the use of open-source implementation of Count-Min Sketches (CMS) in a testbed encompassing a real P4 ASIC switch. We implemented different versions of CMS sketches, suitable for ports at 100 and 400 Gb/s and evaluated its performance in terms of latency introduced to the packets and memory consumption of the P4 ASIC. As observed, every single packet must traverse all the stages in the pipeline of the P4 implementaion, resulting in a total latency that varies between $0.95$ and $1.35~\mu s$ of processing time.


\section*{Acknowledgments}
The authors would like to acknowledge EU H2020 project B5G-OPEN (grant no. 101016663), the Spanish project ENTRUDIT (grant no. TED2021-130118B-I00) and project FuN4Date-Redes Project under Grant PID2022-136684OB-C21 funded by Agencia Estatal de Investigacion of Spain.


%

\bibliographystyle{IEEEtran}
\bibliography{refsp4}


\end{document}